# Hole crystallization in the spin ladder of $Sr_{14}Cu_{24}O_{41}$[*]

P. Abbamonte[a,b], G. Blumberg[c], A. Rusydi[a,d], A. Gozar[c,e], P. G. Evans[f], T. Siegrist[c], L. Venema[d], H. Eisaki[g], E. D. Isaacs[c,h], and G. A. Sawatzky[i]

[a]Brookhaven National Laboratory, Upton, NY, 11973, USA

[b]Department of Physics and Astronomy, SUNY Stony Brook, Stony Brook, NY, 11794, USA

[c]Bell Laboratories, Lucent Technologies, Murray Hill, NJ, 07974, USA

[d]University of Groningen, 9747 AG Groningen, The Netherlands

[e]Department of Physics, University of Illinois at Urbana-Champaign, Urbana, IL, 61801, USA

[f]Department of Materials Science & Engineering, University of Wisconsin, Madison, WI, 53706, USA

[g]Nanoelectronics Research Institute, AIST, 1-1-1 Central 2, Umezono, Tsukuba, Ibaraki, 305-8568, Japan

[h]Center for Nanoscale Materials, Argonne National Laboratory, Argonne, IL 60439, USA

[i]Department of Physics and Astronomy, University of British Columbia, Vancouver, B.C., V6T-1Z1, Canada

**One of the deepest questions in condensed matter physics concerns what other phases compete with superconductivity in high-transition-temperature (high-$T_c$) superconductors. One candidate is the "stripe" phase[1,2,3], in which the carriers (holes) condense into rivers of charge separating regions of antiferromagnetism. A related but lesser known system is the "spin ladder", which consists of two coupled chains of magnetic ions forming an array of rungs. A doped ladder can be thought of as a high-$T_c$ material with lower dimensionality, and has been predicted to exhibit both superconductivity[4,5,6] and an insulating "hole crystal"[4,7,8] phase in which the carriers are localised through many-body interactions. The competition between the two resembles that between stripes and superconductivity in high-$T_c$ materials[9]. Here we report evidence, from resonant x-ray scattering[10], for the existence of a hole crystal in the doped spin ladder of $Sr_{14}Cu_{24}O_{41}$. This phase exists without a detectable distortion in the structural lattice, indicating it arises from many-body effects. Our measurements confirm theoretical predictions[4,7,8] and support the picture that proximity to charge ordered states is a general property of superconductivity in copper-oxides.**

$Sr_{14}Cu_{24}O_{41}$ (SCO) is a layered material consisting of two different types of copper-oxide sheets – a $CuO_2$ "chain" layer and a $Cu_2O_3$ "ladder" layer (see Ref. 11 for a picture). These two sublayers are separated by Sr atoms and stack in an alternating fashion along the $b$ crystallographic direction. The ladders and chains are parallel and run along the $c$ direction, but are structurally incommensurate i.e. the ratio of their lattice parameters $c_L / c_c = \sqrt{2}$ is not a rational number. As a result SCO is internally strained and has a large unit cell with low temperature lattice parameters $a$ = 11.47 Å, $b$ = 13.35 Å, $c$ = 27.3 Å ≈ 7 $c_L$ ≈ 10 $c_c$ [12].

SCO is an intrinsically hole-doped material with 6 hole carriers per formula unit, of which 5.2 reside in the chain layer and 0.8 in the ladder[13]. SCO has the striking property that, when alloyed with Ca and subjected to hydrostatic pressure of 3 GPa, it superconducts with $T_c = 12$ K[14]. Without Ca, however, it exhibits all the transport signatures of a charge density wave (CDW), including a screening mode in impedance measurements[15,16], a pinning mode in microwave conductivity[17], a giant dielectric constant[15,16], and a nonlinear current-voltage (I-V) curve[15], which together indicate that the carrier density is modulated in real space. These observations are typical of conventional Peierls CDW materials like $NbSe_3$ or $K_{0.3}MoO_3$[18] in which the carrier density is modulated by a distortion in the crystal structure, driven by the electron-lattice interaction. However a hole crystal, which is expected to compete with superconductivity in doped ladders and is driven instead by many-body interactions, would bear these same signatures. Could $Sr_{14}Cu_{24}O_{41}$ contain a hole crystal?

Distinguishing between the two requires determining whether the modulation is tied to a lattice distortion or occurs only among the carriers (though perhaps influencing the lattice indirectly). Recently we demonstrated that resonant x-ray scattering[10] at energies near the $K$ shell of oxygen ($1s \rightarrow 2p$ transition) is directly sensitive to hole ordering. In this method the x-ray energy is tuned to the oxygen mobile carrier prepeak (MCP, see Fig. 1) at which scattering from the holes is selectively enhanced by $>10^3$. Here we apply this technique to search for hole ordering in SCO. This material is a particularly interesting case since it has hole carriers in both the ladder and chain layers, and the MCP is split into resolvable ladder and chain features[13]. Each provides a separate enhancement, permitting ordering in two layers to be distinguished.

Single crystals of SCO were grown by travelling solvent floating zone techniques[19], cut to (0,0,1) orientation, polished with diamond film down to 0.1 μm roughness, and annealed in $O_2$ at 120°C to condition the surface. Resonant soft x-ray scattering (RSXS) measurements were carried out on the X1B undulator line at the National Synchrotron Light Source with a 10-axis, ultrahigh vacuum diffractometer.

Here we will use the Miller indices $H$ and $K$ to denote periodicities along the $a$ and $b$ directions, respectively. For the $c$ direction we will respectively use $L_c$, $L_L$, and $L$ to denote periodicity in terms of reciprocal units of the chain, ladder, and total unit cell, i.e. $L = 7L_L = 10L_c$.

Previous studies of SCO with inelastic neutron scattering[20] have reported evidence for a spin dimerization in the chain layer with a periodicity of $5c_c$. This has been corroborated by neutron[12], x-ray[11,21] and electron[22] diffraction, which have shown "superlattice" Bragg reflections that index roughly as $L_c = 2n \pm 0.2$[11,12]. This phenomenon is unlikely to account for the observed CDW transport properties, however, since transport in this system is determined by the ladders[23]. Moreover, in terms of the true unit cell these reflections index as $L = 20n \pm 2$ (always an integer) so are not superlattice reflections in the true sense[12,24]. So far no true superlattice, with a periodicity different from the 27.3 Å unit cell, has been observed in this material.

In Figure 2 we show reciprocal space maps around $(H,K,L_L) = (0,0,0.2)$ at $T = 28$ K, for x-ray energies both off and on the MCP of the ladder. Off-resonance only a specular "rod" is visible, due to reflectance from the sample surface. If tuned to the ladder MCP, however, a pronounced superlattice reflection appears, centred at $(H,K,L_L) = (0,0,0.200 \pm 0.009)$, indicating the existence of a modulation along the ladder with period $5.00 \pm 0.24$ $c_L$. This reflection is commensurate but is truly a superlattice peak, since it occurs at $L = 1.4$ (or $L_c = 0.14$) so does not have the periodicity of the 27.3 Å unit cell. In particular it should not be confused with the chain dimerization reflections, which have a different periodicity.

Our central observation is the extremely unusual energy dependence of this peak. It was tracked through the oxygen $K$ edge where it was found to be visible *only* for incident energies in resonance with the ladder MCP (Fig. 1) – an observation reproduced in two samples from different growth boules. The reflection is undetectable at all other energies, including the O$K$ edge jump, eliminating the possibility that it arises from a distortion in the crystal structure. In x-ray terminology, the peak responds to the anomalous scattering factors of the doped holes, and not those of the oxygen atoms, and therefore indicates a standing wave in the hole density without a (significant) lattice distortion. From its energy dependence and commensurate 5.00 $c_L$ periodicity it is clear this modulation originates in the ladder substructure. The simplest interpretation is a crystallized state of holes in the ladder, which is almost degenerate with superconductivity but wins out under the set of parameters relevant to SCO.

To further characterize this reflection it was tracked through the $L$ edge of copper ($2p \rightarrow 3d$ transition, Fig. 3), where it is also visible and notably still resides at $L_L = 0.2$, verifying that it disperses according to Bragg's law. Scattering at transition metal $L$ edges is known to be sensitive to spin modulations[25,26], however close inspection reveals that it resonates not at the $L_{3/2}$ maximum but at the $L'_{3/2}$ shoulder, which arises from holes on the neighbouring ligands[27]. So the modulation has no obvious magnetic character.

Finally, the x-rays were tuned to the ladder MCP and $L_L$ scans carried out at different temperatures (Fig. 4). The hole modulation is visible below $T_c \approx 250$ K and monotonically increases with cooling. The onset is gradual but close to the $T_c = 210$ K estimated from low frequency dielectric spectroscopy[28], suggesting that the hole crystal is responsible for the CDW signatures in transport. The width of the reflection is temperature-independent even near the transition, so the correlation length is limited by some mechanism other than thermodynamics, perhaps impurities[18] or intrinsic quantum fluctuations.

Our study of $Sr_{14}Cu_{24}O_{41}$ corroborates the prediction by Dagotto[4] of hole crystallization in doped ladders and supports the picture that proximity to charge ordered states is a general property of superconductivity in copper-oxides. RSXS does not permit precise determination of the form factor of the hole crystal, but no harmonic was seen at $L_L = 0.4$ suggesting a sinusoidal, delocalised modulation as discussed by White[7] rather than a fully localized Wigner crystal[29]. The peak width (Fig. 2) shows that the modulation is coherent across ~ 50 neighbouring ladders, demonstrating significant inter-ladder coupling.

Since a hole crystal is charged, the reader may wonder why we do not see a distortion in the lattice which might be induced electrostatically. Such a modulation must exist, but would be of order the amplitude of the hole modulation itself which is likely ~$10^{-2}$ electrons[7]. By contrast, the density modulation of a structural Peierls CDW is of order the atomic number, $Z$. So the scattering power of a hole crystal is nominally weaker by $(10^{-2}/Z)^2 \sim 10^{-6}$. Our point is not that the structural modulation is truly zero, but that electronic correlations, rather than the electron-phonon interaction, drive the transition.

A significant open question concerns the relationship between the observed wavelength of $\lambda = 5.00\ c_L$ and the estimated[13] hole density in the ladder of $\delta = 0.057$ holes / copper. In models of hole crystallization[7,8] $\lambda = 1/\delta\ c_L$ or $2/\delta\ c_L$ in the strong and weak coupling regimes, respectively, which would require $\delta = 0.20$ or $\delta = 0.40$. These models neglect many residual interactions and details of the chemistry, but this relationship is resilient to such corrections. This may indicate a problem with estimates of $\delta$, but it is worth noting that the hole crystal is commensurate with the lattice to within the measurement precision, suggesting that it is partly stabilized by elastic Umklapp processes. These can be significant for a commensurate hole crystal and perhaps strong enough to draw in extra charge from the chains. Another clue lies in the large transverse coherence length, which demonstrates significant inter-ladder interactions, and the relationship between λ and δ for a truly two-dimensional ordering pattern would not be so simple. It is therefore worth extending such models to the case of coupled ladders, or where the ladder interacts with a charge bath with which it may interchange carriers freely.

**Acknowledgements** We gratefully acknowledge J. Grazul and M. Sergent for help with sample polishing and discussions with I. Affleck, J. B. Marston, Y.-J. Kim, P. M. Platzman, J. M. Tranquada, A. Tsvelik, and T. M. Rice. This work was supported by the U.S. Department of Energy, NWO (Dutch Science Foundation), and FOM (Netherlands Organization for Fundamental Research on Matter).

**Competing interests statement:** The authors declare that they have no competing financial interests.

**Correspondence** and requests for materials should be addressed to P.A. (e-mail: abbamonte@bnl.gov).

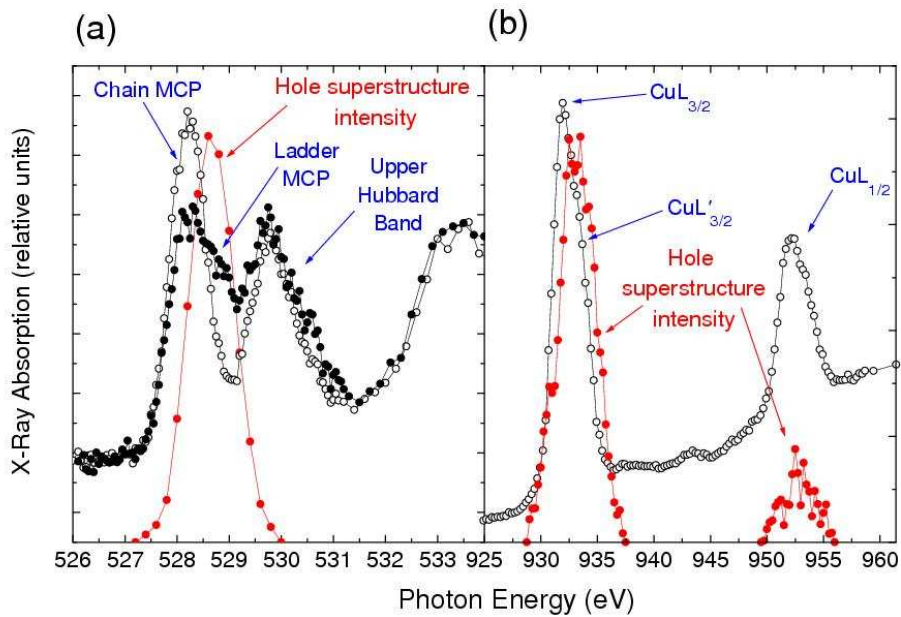

**Figure 1** Energy dependence of the hole superstructure reflection compared to x-ray absorption spectra. (Black symbols) absorption spectra of $Sr_{14}Cu_{24}O_{41}$, taken *in situ* in fluorescence-yield mode, in the vicinity of (a) the oxygen *K* edge, which is a $1s \rightarrow 2p$ transition, and (b) copper $L_{3/2,1/2}$ edges, which are $2p \rightarrow 3d$ transitions where the core hole is left with its spin either parallel ($j = 3/2$) or antiparallel ($j = 1/2$) to its orbital moment. Open circles were taken with the photon polarization **E**||**a** and filled circles with approximately **E**||**c**. The data are in good agreement with Ref. 13. Labels "chain" and "ladder" indicate the respective oxygen mobile carrier prepeaks (MCP) where scattering from the holes is enhanced. (Red symbols) integrated intensity of the hole superstructure reflection as a function of incident photon energy. The reflection is visible only when the x-ray energy is tuned to the ladder MCP or the copper $L_{3/2}'$ ligand hole sideband, indicating the presence of a standing wave in the hole density in the ladder.

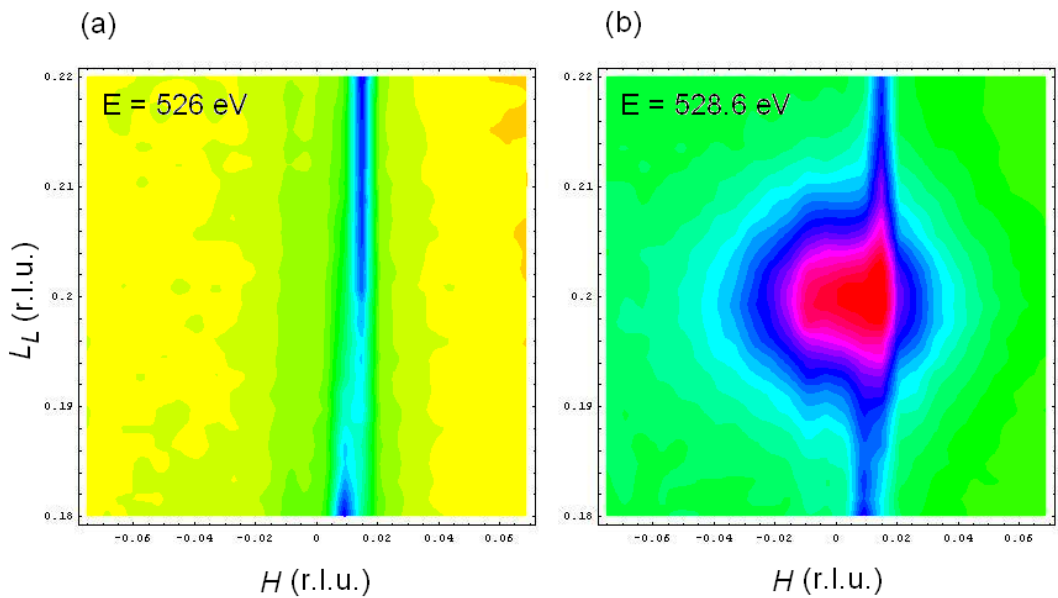

**Figure 2** Appearance of the hole superstructure peak on resonance. (a) Off-resonance ($E$ = 526 eV) reciprocal space map around $(H,K,L_L) = (0,0,0.2)$. The rod at $H$ = 0.01 is the specular reflectance from the surface, which is displaced from $H$ = 0 because of surface miscut. The width of this rod indicates our transverse momentum resolution. (b) Same reciprocal space region with the x-ray energy tuned to the ladder MCP ($E$ = 528.6 eV). A pronounced superlattice reflection appears at $L_L = 0.200 \pm 0.009$, indicating the presence of a commensurate standing wave in the ladder hole density with period $\lambda = 5.00\ c_L$. This reflection indexes to neither the 27.3 Å unit cell nor the previously-reported chain dimerization reflections[11,12,21,22]. The peak width gives longitudinal and transverse coherence lengths of $\xi_c$ = 255 Å = 65.3 $c_L$ and $\xi_a$ = 274 Å = 24.9 $a$, respectively. The hole modulation is registered across 50 neighbouring ladders, indicating significant inter-ladder coupling in this system.

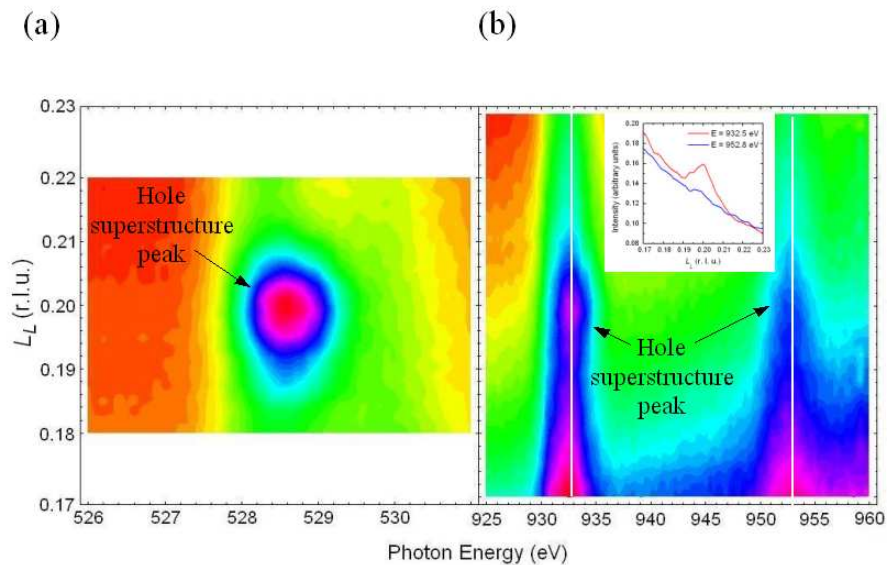

**Figure 3** Energy- and $L_L$-dependence of the hole superstructure reflection. (a) Peak intensity and $L_L$ position across the oxygen $K$ edge and (b) the copper $L_{3/2}$ and $L_{1/2}$ edges. The hole crystal reflection is clearly visible at both the ladder MCP (Bragg angle = 36.1°) and the copper $L_{3/2}$ edge (Bragg angle = 19.9°). In both cases it resides at $L_L$ = 0.200, indicating that, while visible only at select energies, the reflection nonetheless disperses according to Bragg's law. This establishes it as a coherent, bulk phenomenon. The background in (b) is from the specular surface reflection, which is strong at the copper $L$ edge. (Inset) $L_L$ scans on the $L_{3/2}$ (932.5 eV) and $L_{1/2}$ (952.8 eV) peaks, corresponding to the sections indicated with white lines.

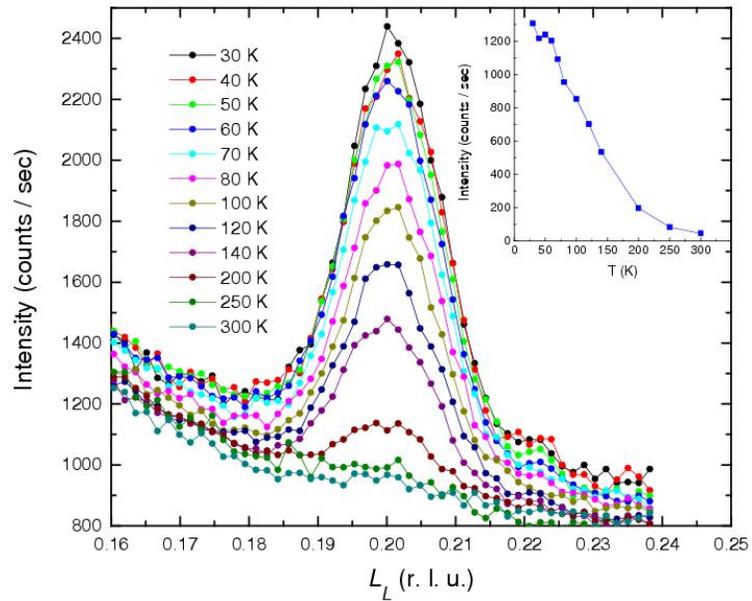

**Figure 4** Temperature dependence of the hole crystal. The superlattice reflection becomes visible below T = 250 K. At its maximum (28 K) the peak count rate is 1500 photons / sec on a fluorescence background of 910 photons / sec. The position and width of the peak are temperature-independent. (Inset) Integrated intensity of the peak as a function of temperature, showing gradual, crossover behaviour, in reasonable agreement with Ref. 28.